# Development of a PCI Express based Readout Electronics for the XPAD3 X-Ray Photon Counting Imager

A. Dawiec, B. Dinkespiler, P. Breugnon, F. Bompard, K. Arnaud, P.-Y. Duval, S. Godiot, S. Hustache, K. Medjoubi, J.-F. Berar, N. Boudet, C. Morel

*Abstract*–XPAD3 is a large surface X-ray photon counting imager with high count rates, large counter dynamics and very fast data readout. Data are readout in parallel by a PCI Express interface using DMA transfer. The readout frame rate of the complete detector comprising 0.5 MPixels amounts to 500 images per second without dead-time.

## I. INTRODUCTION

THE XPAD3 camera is a large surface silicon photon counting detector (12 x 7.5 cm$^2$) developed at CPPM to meet the requirement of experiments on high flux [1] and high brilliance 3$^{rd}$ generation synchrotron sources and on low dose photon counting cone beam (CB) CT [2, 3]. The detector is composed of eight tiled modules operating in parallel in order to obtain a large detection surface with minimal dead areas. Essential features of the detector are its energy thresholds that can be set individually for every pixel and its fast data acquisition system that can read out the complete detector every 2 ms without dead-time. Three complete detectors were assembled at CPPM.

In this paper, we describe the readout architecture of the XPAD3 camera based on a PCI Express interface and present preliminary scans obtained with a monochromatic beam at ESRF and a polychromatic beam from a X-ray tube manufactured by RTW.

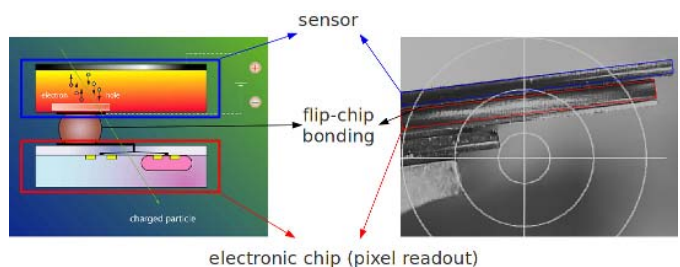

Fig. 1. Scheme of the XPAD3 hybrid pixel detector: a silicon sensor is connected to the electronic readout circuit using flip-chip and bump bonding technologies

## II. XPAD3 HYBRID PIXEL DETECTOR

The XPAD3 detector is a hybrid pixel detector built of eight modules, each composed of seven XPAD3 photon counting circuits [4] connected to a 500 µm thick silicon sensor (Fig. 2). The concept of the XPAD3 detector is shown in Fig. 1. The sensor is connected to its electronic readout circuit using flip-chip and bump bonding technologies. Both the sensor and its readout chips are pixelized with the same pitch and each one of the sensor pixels is connected to a pixel of the electronic readout circuit.

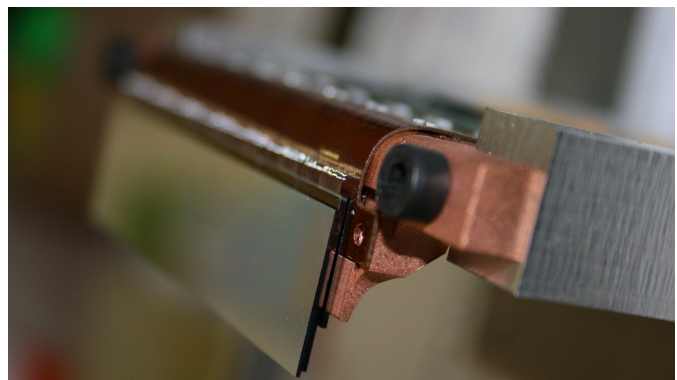

Fig. 2. Picture of a module of the XPAD3 camera made of a pixelized silicon sensor comprising 67'200 pixels connected to seven XPAD3 chips

The XPAD3 chip has been designed in IBM 0.25 µm technology. It is composed of 9'600 pixels organized in 80 columns of 120 rows. The detector specifications are summarized in Table I. Every pixel has a 12 bits counter with overflow. Reading the overflow at a higher rate than the counter actually fills up allow to increase the dynamics of the counter almost without limitations, but the depth of the software encoding.

TABLE I. SPECIFICATIONS OF THE XPAD3 CAMERA

| | |
|---|---|
| Number of pixels | 537'000 |
| Pixel size | 130 x 130 µm$^2$ |
| Counter depth | 12 bits + overflow |
| Readout time | 2 ms |
| Maximum count rate | 10$^6$ photons/pixel/s |
| On the fly readout | Yes |
| Discrimination mode | Single leading edge threshold |
| Minimum threshold | ~ 4 keV |

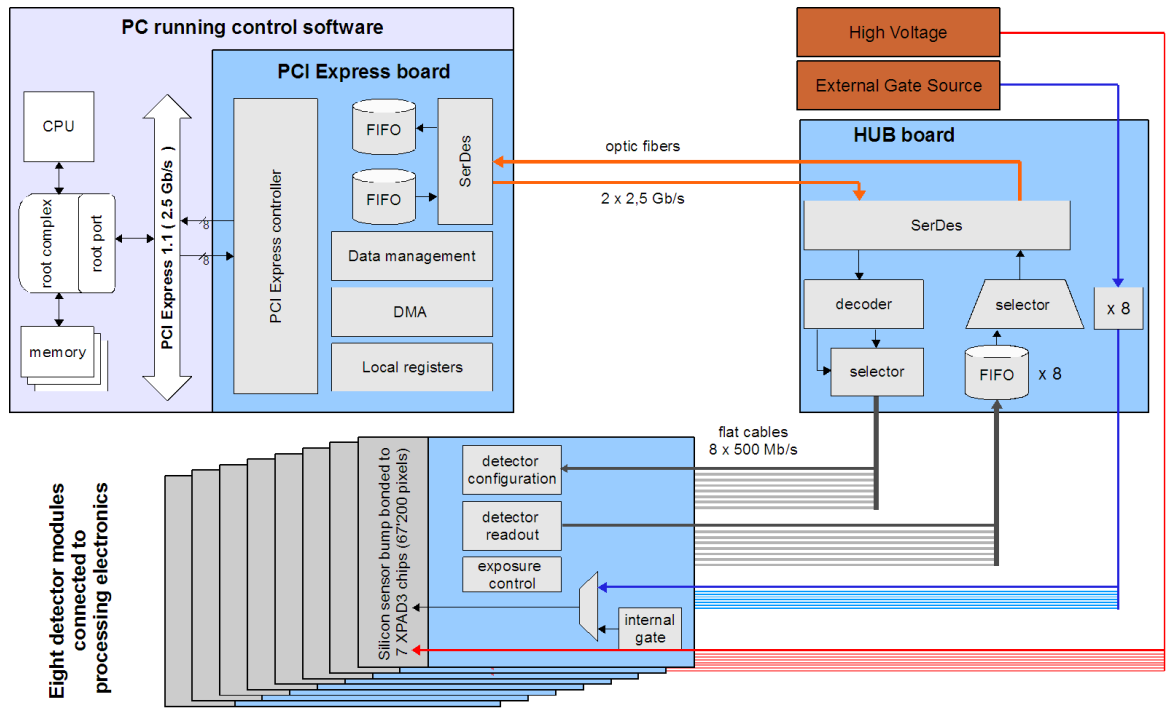

Fig. 3. Architecture of the XPAD3 detector

## III. READOUT ARCHITECTURE OF THE IMAGER

The readout architecture of the imager can be divided into three main blocks: the XPAD3 detector, a HUB board and a PCI Express interface, which is connected to a PC running the data acquisition (DAQ) software. A detector simplified architecture is presented in Fig. 3. Fig. 4 shows a picture of the detector mounted on a goniometer at the synchrotron SOLEIL.

### A. The XPAD3 Detector

Eight independent modules, each of them being connected to its own readout electronics, form the XPAD3 camera. The electronic boards of the modules can configure the XPAD3 circuits with calibration data, process on the fly the overflow bit during exposition, read the counters of the pixels and organize the image data in a format preselected by the user (counts per pixel stored on 2 or 4 bytes). Furthermore, a gate signal may be generated locally in the board or can be provided by an external source. The detector bias voltage and the analog calibration pulses are distributed individually to every electronic board.

### B. The HUB Board

The HUB board stays midway within the detector readout chain. All messages between the DAQ software and the detector modules pass trough the HUB board. Eight modules are connected to this board via a set of flat copper cables. Communication with the PC is done trough a pair of optical fibers. The data from the DAQ are first decoded according to the addresses encoded in the data frames. Each one of the detector modules writes its data to dedicated buffers implemented on the HUB board. The HUB selector then sequentially extracts and sends them to the PCI Express interface via two optical fibers. Besides detector data management functions, the HUB board can operate communication diagnostic tests and distribute external gate signals to every module in order to perform time resolved experiments.

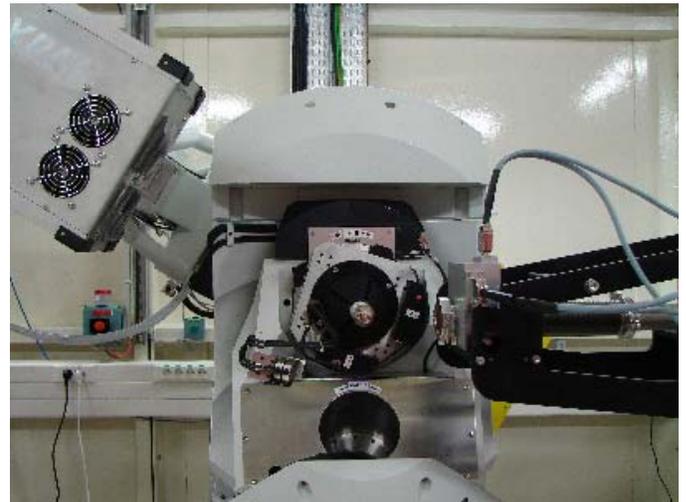

Fig. 4. Picture of the XPAD3 camera mounted on a goniometer at the synchrotron SOLEIL.

### C. PC with PCI Express Board

In order to read out the detector as fast as possible, a PCI Express interface is used to transfer the data arriving from the HUB board directly to the computer embedded memory. A PCI Express x8 (8 lanes) in 1.1 standard is used. It has a theoretical bandwidth of 32 Gbits/s of un-encoded data (before

8b/10b encoding). In order to fully benefit from the high bandwidth of the PCI Express link, we use direct memory access (DMA) to transfer data from the PCI Express interface to the PC's memory. Table II summarizes measured data rates over PCI Express 1.1 x8 during DMA data transfer.

TABLE II. PCI-E MEASURED DATA RATES

| | |
|---|---|
| Writing (DAQ to PCI-E): | 500 MBytes/s |
| | 500 images/s |
| Reading (PCI-E to DAQ): | 1000 MBytes/s |
| | 1000 images/s |

The PCI-E board includes a FPGA chip connected to several SFP optical transceivers. The FPGA embedded system comprises a PCI Express controller, a DMA engines (reading and writing) and a data management unit. The DMA transfer configurations (address and size) are stored in a set of local registers that can be dynamically changed by the software application. The data from the optical fibers are first de-serialized and stored in the reception buffer. Then the DMA transfer is stored to the address specified in the register as soon as the buffer has received the data size set by the user. The implemented setup allows to send data from the complete detector to the PC in 2 ms.

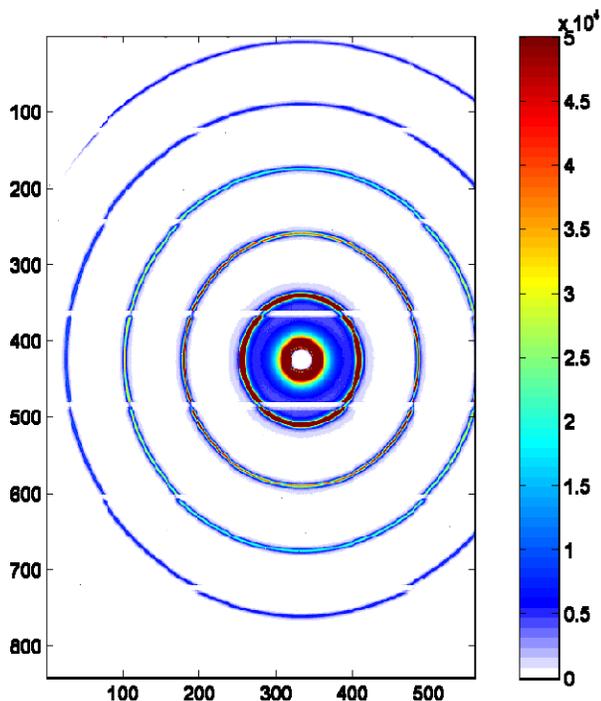

Fig. 5. Diffraction rings on a powder of silver behenate (AgBe) recorded at ESRF using a monochromatic beam of 16 keV X-rays

Today the size of the RAM embedded in the PC limits the size of the image sequence that can be acquired. In the future, compression, fast disk storage or network distribution could release this limitation.

## IV. PRELIMINARY SCANS

The specification of the XPAD3 camera, its programmable thresholds, its high count rate dynamics and fast data readout makes it a suitable device for experiments in the fields of crystallography and small animal imaging capable of capturing dynamic processes.

In Fig. 5, the result of powder diffraction is presented. The thresholds of the detector were set at 8 keV and the detector was exposed to a 16 keV beam over 100 s. Debye rings resulting from the diffraction of the X-rays on a silver behenate (AgBe) powder are clearly visible.

As an illustration, Fig. 6 shows the result of a CB CT scan of a gecko acquired with the XPAD3 camera using a RTW Mo target X-ray tube operated at 40 kV [5]. In order to obtain a pseudo-monochromatic beam strongly peaked around 17 keV, a Nb/Mo filter was inserted in between the X-ray tube and the object under study.

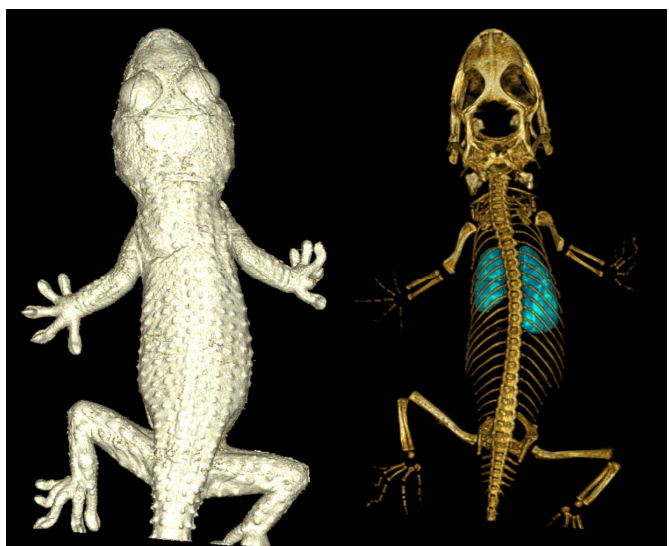

Fig. 6 Volume renderings of a CB CT scan of a gecko.

## V. CONCLUSION

The architecture of the XPAD3 detector and the results obtained so far with this device open the possibility to make series of experiments which were hardly feasible up to now. Moreover parallel readout of the detection modules makes it possible to scale it up to larger detectors without loss in image acquisition speed.